\documentclass[final,letter,5p,times,twocolumn]{elsarticle}
\usepackage{lineno,hyperref}
\biboptions{sort&compress}
\journal{Elsevier}

\usepackage[T1]{fontenc}
\usepackage[latin9]{inputenc}
\usepackage{amsmath}
\usepackage{graphicx}

\usepackage{color}
\usepackage{graphicx}
\usepackage{multirow}
\usepackage{amsmath,bm}

\newcommand {\snn}	{\sqrt{s_{_{\rm NN}}}}
\newcommand {\gevc}	{GeV/$c$}

\newcommand {\Nch}	{N_{\rm ch}}

\newcommand {\Zr}	{$^{96}$Zr}
\newcommand {\Ru}	{$^{96}$Ru}
\newcommand {\RuRu}	{$^{96}_{44}$Ru+$^{96}_{44}$Ru}
\newcommand {\ZrZr}	{$^{96}_{40}$Zr+$^{96}_{40}$Zr}

\newcommand {\pt}	{p_{\rm T}}

\newcommand {\vtt}	{v_{2}\{2\}}
\newcommand {\vtf}	{v_{2}\{4\}}
\newcommand {\vft}	{v_{4}\{2\}}
\newcommand {\phitf} {\mean{\cos{4(\Phi_2-\Phi_{4})}}}

\newcommand{\ac}{{\rm ac}_{2}\{3\}}
\newcommand{\nac}{{\rm nac}_{2}\{3\}}

\newcommand{\betaZr} {\beta_{\rm 3,Zr}}
\newcommand{\betaRu} {\beta_{\rm 2,Ru}}

\newcommand {\mean}[1]	{\langle #1\rangle}

\begin{document}
\begin{frontmatter}	

\title{Probing the nuclear deformation with three-particle asymmetric cumulant in RHIC isobar runs}

\author[1,2]{Shujun Zhao}
\author[2]{Hao-jie Xu\corref{mycorrespondingauthor}}\ead{haojiexu@zjhu.edu.cn}
\author[1,3,4]{Yu-Xin Liu\corref{mycorrespondingauthor}}\ead{yxliu@pku.edu.cn}
\author[1,3,4]{Huichao Song\corref{mycorrespondingauthor}}\ead{huichaosong@pku.edu.cn}

\address[1]{Department of Physics and State Key Laboratory of Nuclear Physics and Technology, Peking University, Beijing 100871, China}
\address[2]{School of Science, Huzhou University, Huzhou, Zhejiang 313000, China}
\address[3]{Collaborative Innovation Center of Quantum Matter, Beijing 100871, China}
\address[4]{Center for High Energy Physics, Peking University, Beijing 100871, China}

\cortext[mycorrespondingauthor]{Corresponding author}

\date{\today}

\begin{abstract}
\RuRu\ and \ZrZr\ collisions at $\snn=200$ GeV provide unique opportunities to study the geometry and fluctuations raised from the deformation of the colliding nuclei. Using iEBE-VISHNU hybrid model, we predict $\ac$ ratios between these two collision systems and demonstrate that the ratios of $\ac$, as well as the ratios of the involving flow harmonics and event-plane correlations, are sensitive to quadrupole and octupole deformations, which could provide strong constrains on the shape differences between \Ru\ and \Zr. We also study the nonlinear response coefficients $\chi_{4,22}$, which show insensitivity to the deformation effect.
\end{abstract}

\end{frontmatter}

\section{Introduction}
Anisotropic flow observed in heavy-ion collisions at Relativistic Heavy-Ion Collider (RHIC) and Large Hadron Collider (LHC) indicate that the created  quark-gluon-plasma (QGP) is a strongly coupled system with small specific shear viscosity~\cite{Adams:2005dq,Adcox:2004mh,ALICE:2010suc,Romatschke:2007mq,Teaney:2009qa,Song:2010mg,Niemi:2011ix,Heinz:2013th,Song:2017wtw}.
Hydrodynamic simulations have successfully described the collective expansion of the QGP fireball and studied various flow observables at RHIC and the LHC~\cite{Gyulassy:2003mc,Kovtun:2004de,Schenke:2010rr,Song:2012ua,
Gale:2012rq,Gale:2013da,Xu:2016hmp,Bernhard:2016tnd,McDonald:2016vlt,Zhao:2017yhj}.
After the hydrodynamic evolution, the initial stage geometry and fluctuations are  translated  into final stage correlations described by various flow observables such as different order flow harmonics, correlations between flow harmonics, event-plane correlations, etc.~\cite{Bilandzic:2010jr, Bilandzic:2013kga, Bhalerao:2013ina, Aad:2014lta,Yan:2015jma, Jia:2017hbm, Zhao:2017yhj,Zhu:2016puf,ATLAS:2018ngv,Zhang:2018lls,Li:2021nas}. On the other hand, these flow observables raised from the collective expansion also depend on the properties of the QGP and the details of the initial profiles. The RHIC isobar runs with \ZrZr\  and \RuRu\ collisions at $\snn=200$ GeV
provide unique opportunities to probe the nuclear structure of the colliding nuclei from the initial stage, since
the uncertainties from the bulk properties of the QGP can be largely reduced through the observable ratios between the two collision systems~\cite{STAR:2021mii,Xu:2021uar}.

The \ZrZr\  and \RuRu\ collisions at $\snn=200$ GeV originally  aimed to search the chiral magnetic effect (CME).  The observed differences in multiplicity distribution ($\Nch$) and anisotropic flow  harmonics  between these two systems ruin the premise that isobar collisions can help identify the CME with enough precision~\cite{Xu:2017zcn,Li:2018oec}, but provide a novel way to constrain the nuclear deformation from heavy ion collisions~\cite{Li:2019kkh}. For the typical initial profile construction for A+A collisions,  the Woods-Saxon distribution for the nuclear density is written as:
\begin{eqnarray}
	\rho &=& \frac{\rho_{0}}{1+\exp{\left[(r-R)/a\right]}} \,, \label{eq:WS} \\
	R &=& R_{0}\left(1 + \beta_{2}Y_{2}^{0}(\theta,\phi) + \beta_{3}Y_{3}^{0}(\theta,\phi)+\cdots\right)\,,
\end{eqnarray}
where $a$ is the diffuseness parameter, $R_{0}$ is the radius parameter, and  $\beta_{2}$ ($\beta_{3}$) is the parameter for quadrupole (octupole) deformation. In the central collisions with impact parameter $b=0$ fm, the deformed nuclei can naturally contribute geometric anisotropy of the overlap area, leading to larger anisotropic flow than the one from spherical nuclei collisions~\cite{Rosenhauer:1986tn, Bernhard:2019bmu, Filip:2009zz, Giacalone:2019pca}.  For the central isobar collisions at RHIC, the observed difference for elliptic flow $v_{2}$  and triangle flow $v_{3}$ between the two collision systems indicate a larger quadrupole deformation for \Ru\ and a larger octupole deformation for \Zr~\cite{STAR:2021mii, Zhang:2021kxj}. In non-central collisions, $v_{2}$ also depends on the diffuseness parameter $a$~\cite{Shou:2014eya}, and the non-trivial bump structure of $v_{2}$ ratio as a function of centrality indicates a thick halo-type neutron skin thickness for \Zr~\cite{STAR:2021mii}, consisting with the predictions from energy density functional theory (DFT)~\cite{Xu:2017zcn, Xu:2021vpn}.
Recently, more observables have been proposed to probe the nuclear structure in relativistic isobar collisions, such as the differences in multiplicity distribution~\cite{Li:2019kkh}, net charge number ($\Delta Q$)~\cite{Xu:2021qjw}, mean transverse momentum ($\mean{\pt}$)~\cite{Xu:2021uar}, $\mean{\pt}$ fluctuations~\cite{Jia:2021qyu}, and spectator neutrons~\cite{Liu:2022kvz}, etc. Benefiting from huge statistics and the strategy to reduce systematic uncertainties in experiment~\cite{STAR:2021mii}, the isobar collisions are expected to provide more precise measurements of neutron skin thickness and nuclear deformations.

Compared with lower order flow harmonics, higher order flow harmonics, as well as the correlations between different flow harmonics are expected to be sensitive  to the initial state deformations~\cite{Yan:2015jma}, which provide more information for precisely probing the nuclear structure differences between the two isobar nuclei.
The flow harmonics and their correlations can be calculated by the multi-particle azimuthal correlations~\cite{Bilandzic:2010jr, Bilandzic:2013kga}:
 \begin{equation}
\mean{m}_{n_1,n_2,...,n_m} \equiv \mean{e^{i(n_{1}\varphi_{k_{1}}+n_{2}\varphi_{k_{2}}+...+n_{m}\varphi_{k_{m}})}},
\end{equation}
Here $\mean{...}$ denotes the sum of all particles of interest (POI) in a given event. The three-particle asymmetric cumulant can be calculated with~\cite{Aad:2014lta,Yan:2015jma, Jia:2017hbm, ATLAS:2018ngv,Zhang:2018lls}:
\begin{equation}
\ac  \equiv \mean{\mean{3} _{2,2,-4}} = \mean{\mean{e^{i(2\varphi_{1}+2\varphi_{2}-4\varphi_{3}})}}
 \end{equation}
Here $\mean{\mean{..}}$ indicates the average of $\mean{..}$ over an ensemble of events. $\ac$ is sensitive to the flow magnitudes and event-plane correlations. It is also directly related to the nonlinear response between the second and fourth order flow vector, which can also be used to extract the corresponding nonlinear response coefficients~\cite{Yan:2015jma}.

We will show in this work that the $\ac$ ratios in isobar collisions are very sensitive to the deformation of the colliding nuclei.
In the absence of non-flow effects, the $\ac$ can be written as~\cite{Yan:2015jma}
\begin{equation}
\ac = \mean{v_{2}^{2}v_{4}\cos{4(\Phi_2-\Phi_4)}},
\end{equation}
 where $\Phi$ is the event-plane of the related flow harmonic.
Partly inherited from $v_2$, $\ac$ is sensitive to the deformation of the colliding nuclei~\cite{STAR:2021mii, Xu:2021vpn, Zhang:2021kxj}.
We will show that the normalized asymmetric cumulant
 \begin{equation}
 \nac \equiv \frac{\ac}{\sqrt{(2\vtt^{4}-\vtf^{4})\vft^{2}}} \label{eq:nac}
 \end{equation}
is also sensitive to the nuclear deformation, even if the contributions from single flow harmonics have been scaled out. Here,
 \begin{eqnarray}
 \vtt^{2} &=& \mean{\mean{2}_{2,-2}},\nonumber \\
  \vft^{2} &=& \mean{\mean{2}_{4,-4}},\nonumber \\
 \vtf^{4} &=& 2\vtt^{4} - \mean{\mean{4}_{2,2-2,-2}} .
 \end{eqnarray}

\begin{table}
	\caption{The iEBE-VISHNU parameters are roughly tuned to reproduce the multiplicity and flow observables reported by the STAR collaboration~\cite{STAR:2021mii}. A detailed description of the following parameters can be found in Ref~\cite{Moreland:2018gsh}.
	\label{tab:VISHNU}}
	\centering{}%
    \begin{tabular}
    {p{2cm}p{1.6cm}p{2cm}p{1.6cm}}
    \hline
	        \multicolumn{2}{c}{Initial condition/Preeq.} & \multicolumn{2}{c}{QGP medium}  \\
    \hline
	    ${\rm Norm}$           & $6.6$ GeV      & $(\eta/s)_{\rm min}$           & $0.134$   \\
	    $p$                   & $0.0$          & $(\eta/s)_{\rm slope}$           & $1.6$ GeV$^{-1}$    \\
	    $\sigma_{\rm flut}$   & $0.91$       & $(\eta/s)_{\rm crv}$            & $-0.29$    \\
	    $r_{\rm cp}$          & $0.88$ fm     & $(\zeta/s)_{\rm max}$         & $0.052$   \\
	    $n_{c} $             & $6.0$          & $(\zeta/s)_{\rm width}$       & $0.024$ GeV   \\
	    $w_{c}$              & $0.36$ fm     & $(\zeta/s)_{T_{\rm peak}}$    & $0.175$ GeV   \\
	    $d_{\rm min}$        & $0.4$ fm       & $T_{\rm switch}$                & $0.151$ GeV   \\
	    $\tau_{\rm fs}$        & $0.37$ fm/$c$ &                                             &    \\
	\hline
	\end{tabular}
\end{table}

\begin{table}
	\caption{WS parameterizations (radius parameter $R_0$, diffuseness parameter $a$, and deformation parameters $\beta_{2}$, $\beta_{3}$) of the \Ru\ and \Zr\ nuclear density distributions with different nuclear deformations, followed by the procedure given in Ref.~\cite{Xu:2021qjw}. The quoted values for $R_0$ and $a$ are in fm. \label{tab:WSDFT}}
	\centering{}%
    \begin{tabular}
    {p{1.5cm}p{1.4cm}p{1.4cm}p{1.4cm}p{1.4cm}}
    \hline
	       &  $\beta_{2}$ &   $\beta_{3}$  &$R_{0}$    & $a$  \\
    \hline
	    Ru-para-I  & 0.12 & 0.00 & 5.093 & 0.478   \\
	    Ru-para-II  & 0.16 & 0.00 & 5.093 & 0.471   \\
	    Zr-para-I   & 0.00 & 0.16 & 5.021 & 0.524   \\
	    Zr-para-II   & 0.00 & 0.20 & 5.021 & 0.517   \\
	\hline
	\end{tabular}
\end{table}

\begin{figure}[!hbt]
	\begin{centering}
		\includegraphics[scale=0.35]{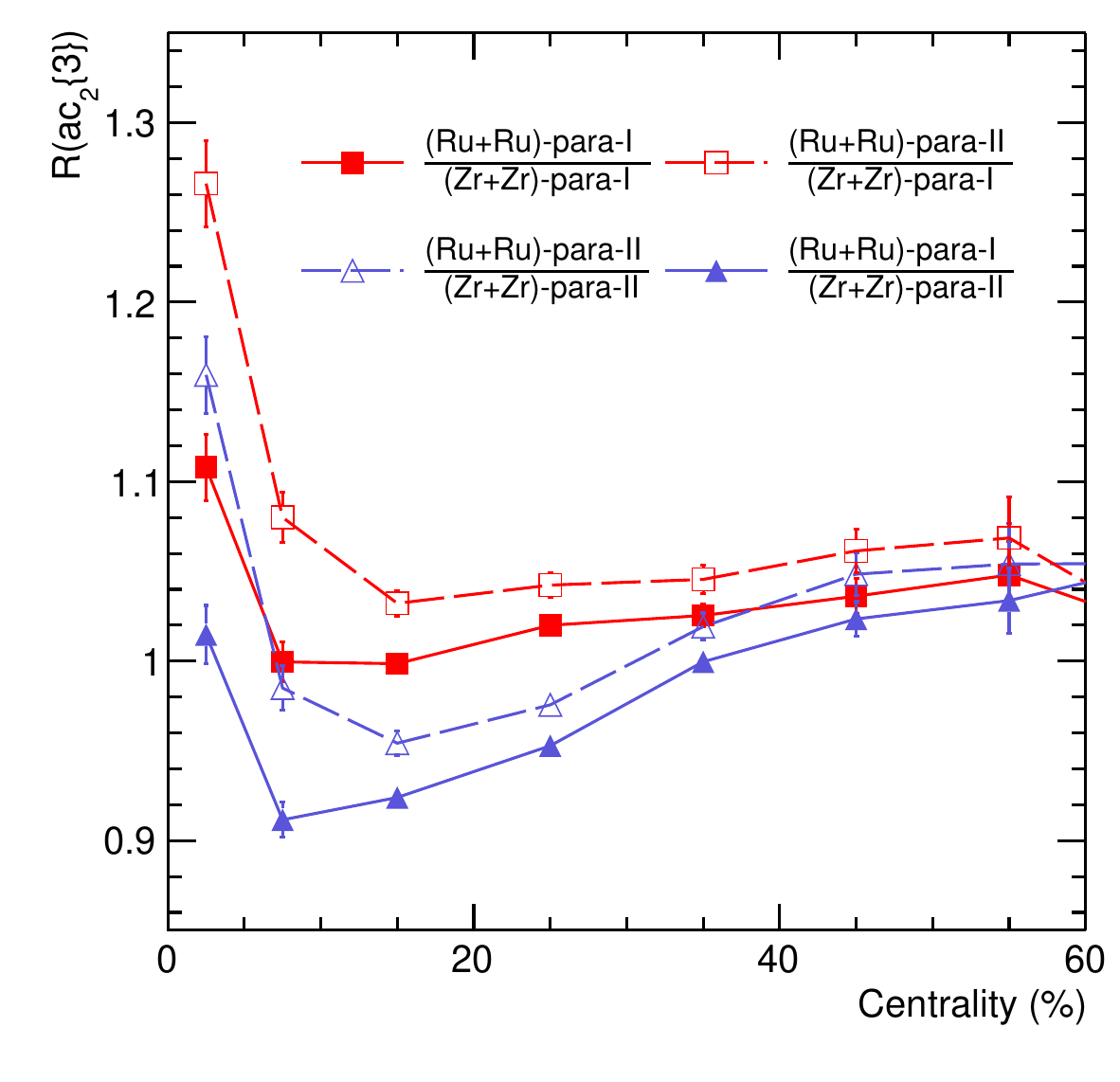}
	\par\end{centering}
	\caption{The centrality dependent $R(\ac)$  in  relativistic isobar collisions, obtained from iEBE-VISHNU simulations. The standard Q-cumulant method is used for charged particles  with $0.2<\pt<2$ \gevc\ and $|\eta|<2$.  \label{fig:ac}}
\end{figure}

\begin{figure*}
	\begin{centering}
		\includegraphics[scale=0.29]{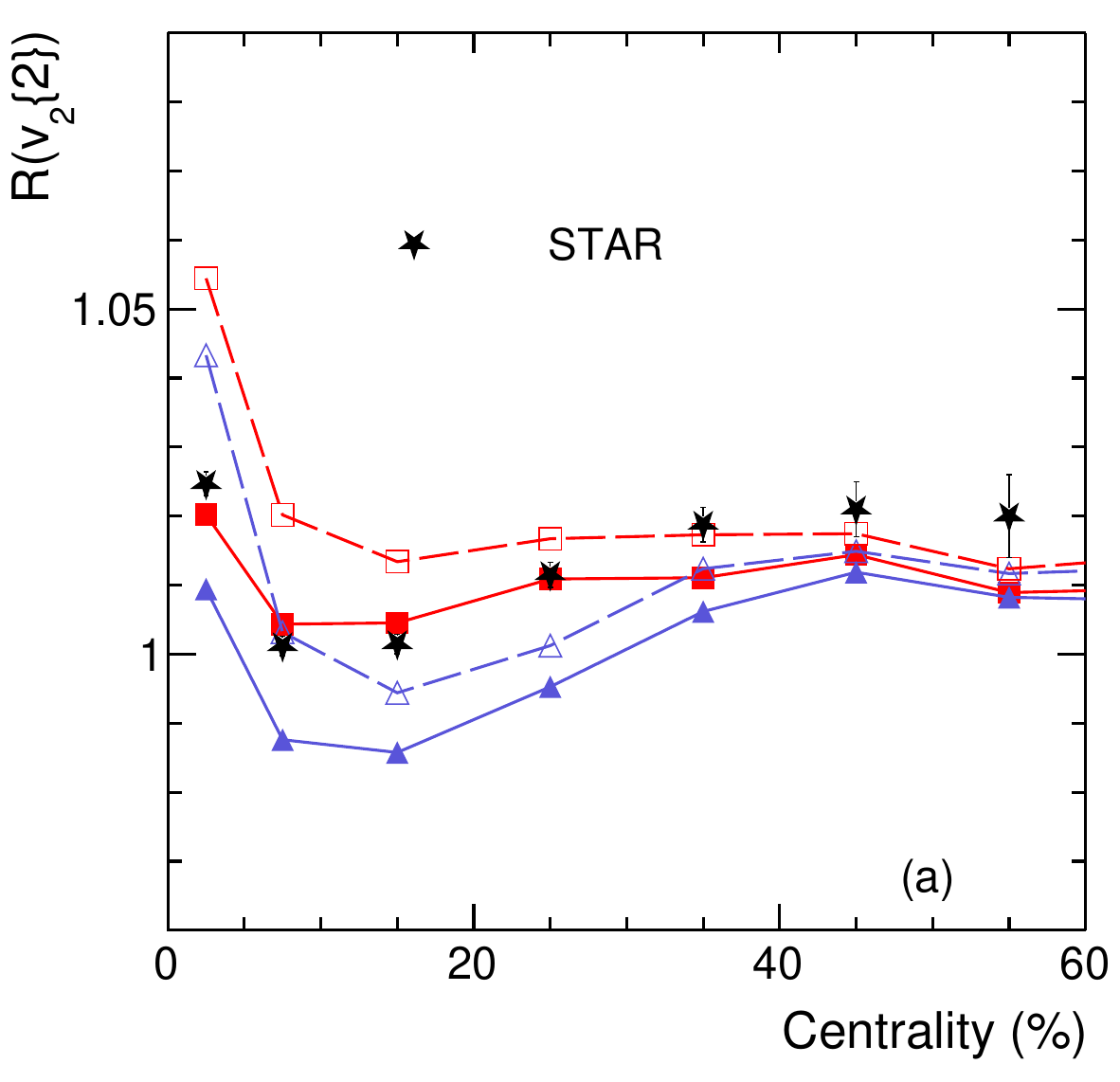}
		\includegraphics[scale=0.29]{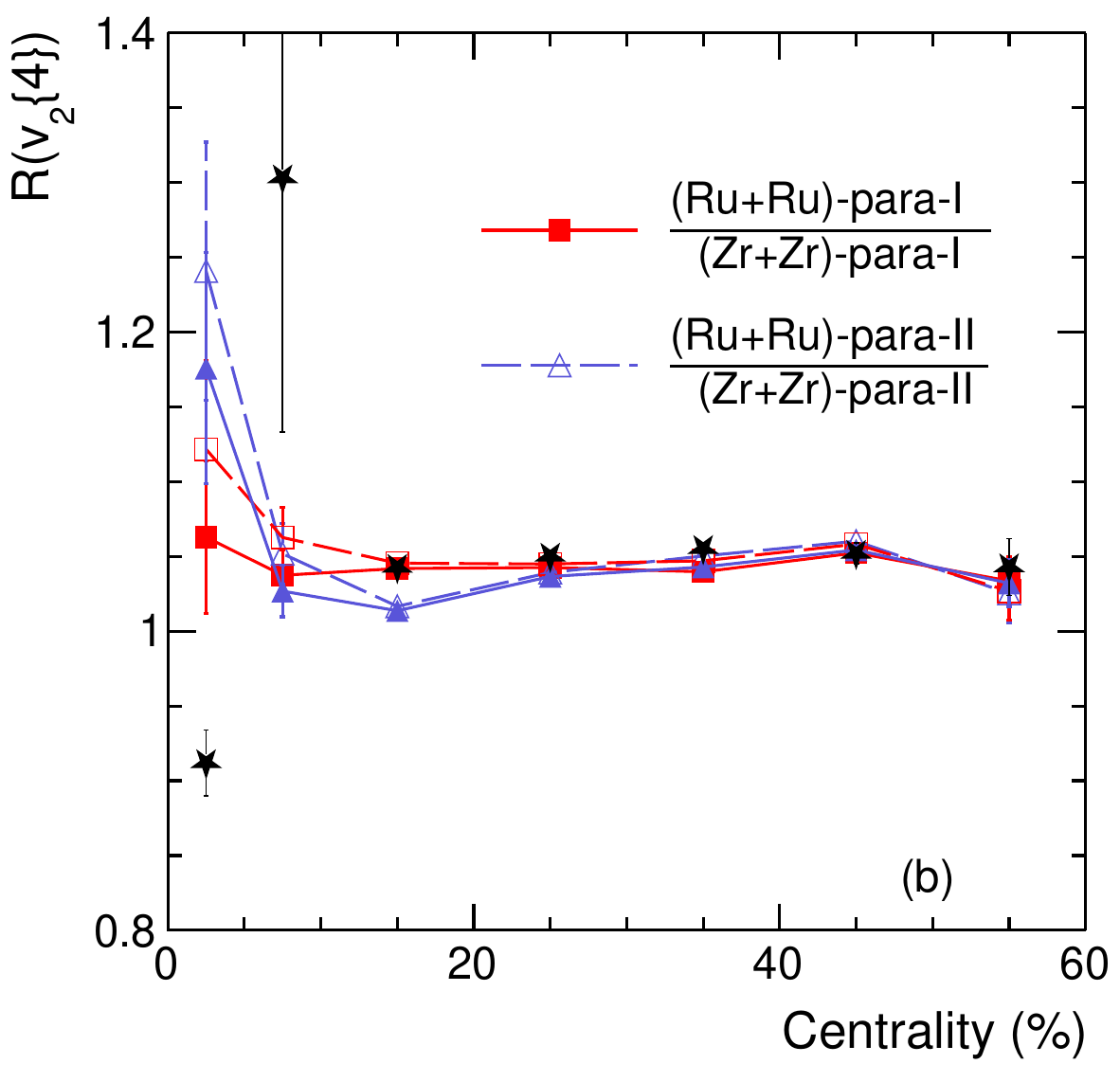}
		\includegraphics[scale=0.29]{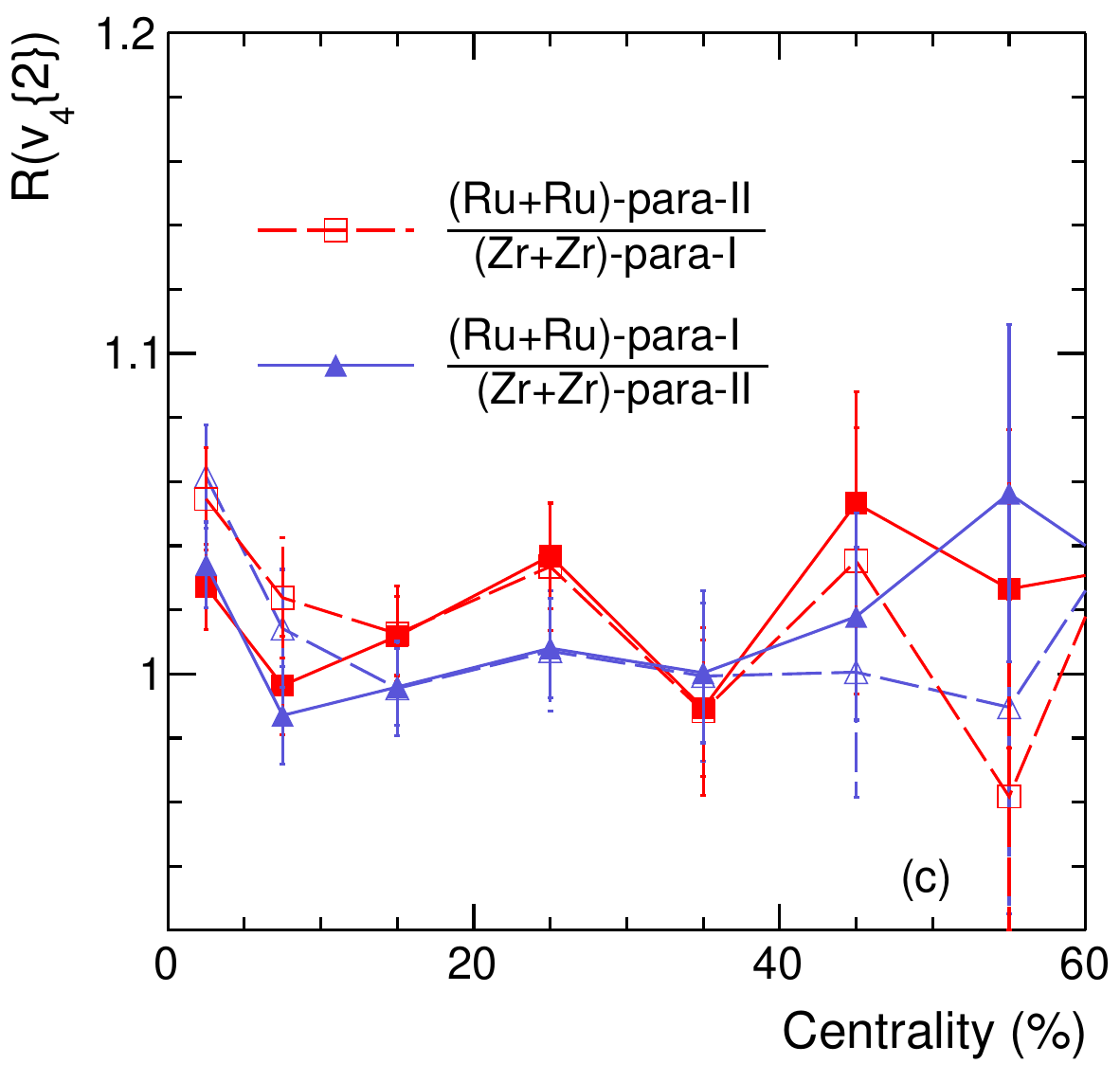}
	\par\end{centering}
	\caption{The centrality dependent (a) $R(\vtt)$, (b) $R(\vtf)$ and (c) $R(\vft$) calculated by iEBE-VISHNU model with different sets of nuclear deformations. The flow harmonics are calculated by the standard Q-cumulant method with $0.2<\pt<2$ \gevc\ and $|\eta|<2$. The data are taken from~\cite{STAR:2021mii}. \label{fig:flow}}
\end{figure*}

In this letter, we will implement iEBE-VISHNU hybrid model to calculate the flow observables and demonstrate that the nuclear deformation not only influences the magnitude of anisotropic flow but also their correlations, which can be  reflected by  $\ac$ and $\nac$ correlations.
Specifically, the effect  on flow harmonics is amplified in  the ratio of $\ac$, and the residual effect from even-plane correlation is reflected by the ratio of $\nac$,  where the ratio is defined as:
\begin{equation}
R(X) \equiv \frac{X_{\rm RuRu}}{X_{\rm ZrZr}}.
\end{equation}

\section{The model}
In this paper,  we implement iEBE-VISHNU to calculate the asymmetric cumulant $\ac$ and the related flow observables in relativistic isobar collisions at $\snn=200$ GeV. iEBE-VISHNU~\cite{Shen:2014vra,Song:2010aq} is an event-by-event hybrid model  that combines (2+1) dimensional viscous hydrodynamics~\cite{Song:2007ux,Song:2007fn} to describe the expansion of the QGP and the hadron cascade model (UrQMD) to simulate the evolution of the subsequent hadronic matter~\cite{Bass:1998ca,Bleicher:1999xi}.
The initial condition of the collision is simulated by the Trento model~\cite{Moreland:2014oya,Bernhard:2016tnd} with the given nuclear density distribution described by Eq.~(\ref{eq:WS}). The parameters for the iEBE-VISHNU simulation are listed in Tab.~\ref{tab:VISHNU}, which are tuned to roughly reproduce the multiplicity and flow observables measured in experiment~\cite{STAR:2021mii}. A more detailed description of those parameters can be found in Ref~\cite{Moreland:2018gsh}.

The nuclear densities with deformation for \Ru\ and \Zr\ have been obtained in Ref.~\cite{Xu:2021uar}, using DFT calculations with the slope parameter of symmetry energy $L(\rho_{c})=47.3$ MeV.
The previous experiments about nuclear structure indicate that $\betaRu =0.16$ with negligible octupole deformation and $\betaZr=0.20$ with negligible quadrupole deformation. However, recent hydrodynamic simulations on relativistic isobar collisions indicate that those values are overestimated~\cite{Nijs:2021kvn}. We therefore choose another set of deformation factors for both \Ru\ and \Zr, i.e. $\betaRu=0.12$ and $\betaZr=0.16$. We found that this can give a better description of elliptic flow ratio $R(v_{2})$ and triangle flow ratio $R(v_{3})$, although the motivation of this study is not focus on quantitative prediction of flow ratios. The Woods-Saxon parameter sets for different deformation factors are listed in Tab.~ \ref{tab:WSDFT}

In this study, the azimuthal correlations are calculated using the standard Q-cumulant method~\cite{Bilandzic:2010jr}
with $Q_{n}\equiv \mean{e^{in\varphi}}$.
All charged particles with $0.2<\pt<2 $ \gevc\ are used.
To reduce statistical uncertainties,
an $|\eta|<2$ pseudorapidity cut is used. iEBE-VISHNU simulations contain part of the non-flow effect from resonance decays. While  the standard Q-cumulant method can not fully reduce the non-flow effect, especially for the $\ac$ which has a large non-flow subtraction method dependence~\cite{ATLAS:2018ngv,Zhang:2018lls}.
The large pseudorapidity cut used in this study can reduce the non-flow contributions to some extend, and we will discuss this in the next section.

\section{Results and discussions}
Figure~\ref{fig:ac} shows the ratio of $\ac$, $R(\ac)$, as a function of centrality in isobar collisions at $\sqrt{s_{NN}}$=200 GeV, calculated from iEBE-VISHNU model.
The comparison of R($\ac$) at most central collisions with different combination of the deformation parameters $\beta_2$ and $\beta_3$ demonstrates that $R(\ac)$ is sensitive to the nuclear deformation.
The trend is similar to the one of $R(\vtt)$ as shown in Fig.~\ref{fig:flow}(a), which decreases from most central to semi-central collision and then increases from semi-central to peripheral collisions. The large $R(\ac)$ and $R(\vtt)$ in the most central collisions are mostly due to the large quadrupole deformation in \Ru~\cite{Zhang:2021kxj}, while
the enhancement trend from semi-central to peripheral collisions is due to the thick halo-type neutron skin in \Zr~\cite{Xu:2021vpn}.
The octupole deformation in \Zr\ give some contributions to $R(\ac)$ and $R(\vtt)$ in the most central collision, and lead to the valley structure in semi-central collisions~\cite{Zhang:2021kxj}.
Compared with $R(\vtt)$, $R(\ac)$ is more sensitive to nuclear deformation and contains more information on flow fluctuations and correlations. Description of their sensitivities on the nuclear structure at a quantitative level can help us to precisely constrain the nuclear deformation factors in isobar collisions. Note that the data of  $R(\vtt)$ in Fig.2 (a) prefer  $\beta_{2,Ru}=0.12$ and $\beta_{3,Zr}=0.16$. While, we should also emphasis that this paper is not aimed to precisely describe the flow data in isobar collisions. More sophisticated extractions of the deformation parameters will be given in the following study~\cite{Zhao:2022prep}.

\begin{figure*}[!hbt]
	\begin{centering}
		\includegraphics[scale=0.32]{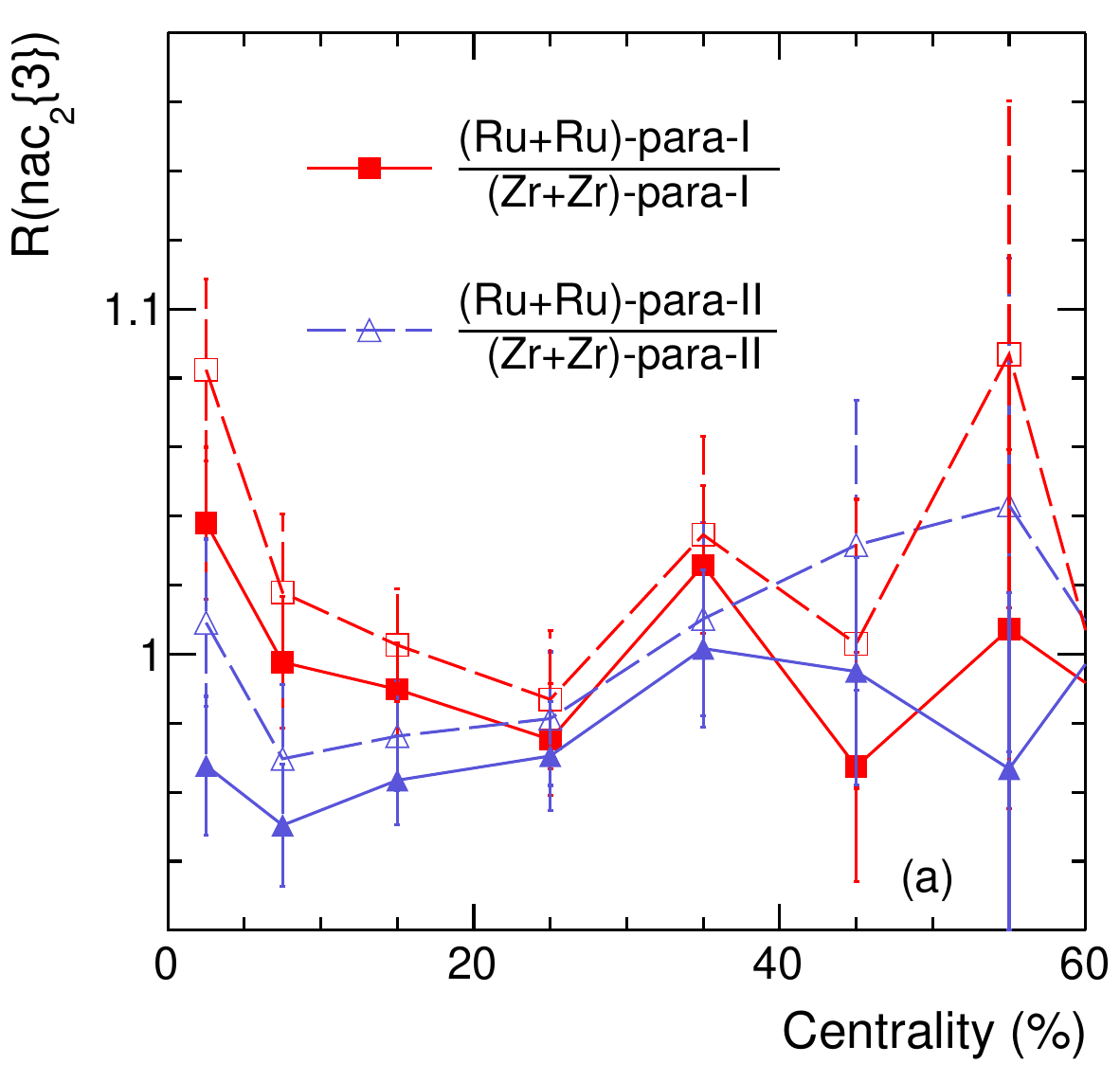}
		\includegraphics[scale=0.32]{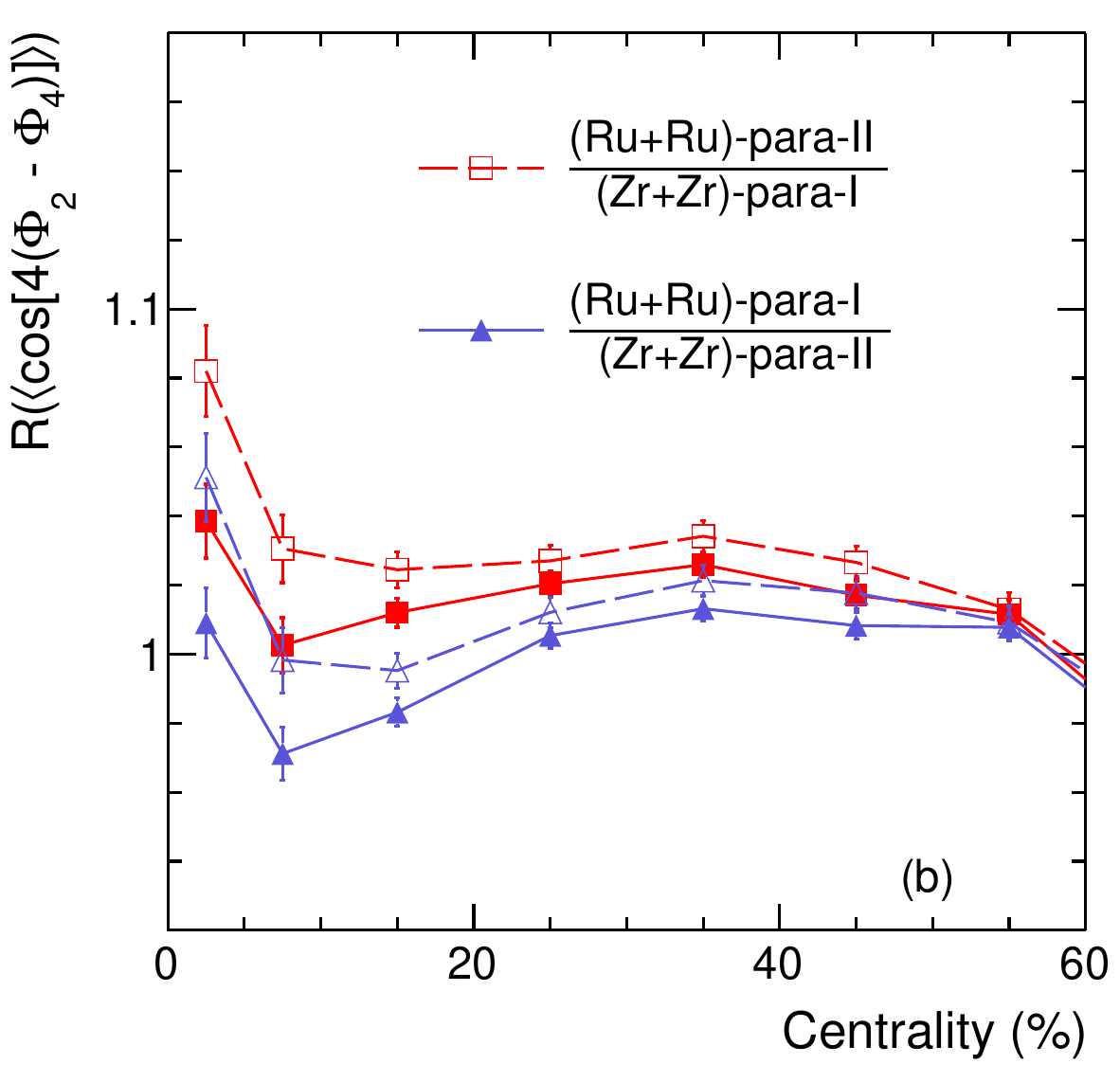}
	\par\end{centering}
	\caption{The centrality dependent (a) $R(\nac)$, (b) $R(\phitf)$ calculated by iEBE-VISHNU model with different sets of nuclear deformations. The observables are calculated by the standard Q-cumulant method with $0.2<\pt<2$ \gevc\ and $|\eta|<2$.\label{fig:nac}}
\end{figure*}

Besides $\vtt$, $\vtf$ and $\vft$ also contribute to $\ac$.
Due to flow fluctuations, $\vtt$  from the two-particle correlation  is larger than $\vtf$ from the four-particle correlation for different collision systems. However, as shown in Fig.~\ref{fig:flow}(a) and (b),
the ratios $R(\vtt)$ and $R(\vtf)$ in isobar collisions present opposite behavior as observed in experiment~\cite{STAR:2021mii}, which indicates the importance of initial state deformation and  fluctuations~\cite{Wang:2022prep}. Note that the $R(\vtf)$ and $R(\vft)$ in the most central collisions also depend on the deformation.
We observe  $R(\vtf)$  deviates  from unity in the most central isobar collisions, while firm conclusion needs high statistical runs.
We note that, comparing to high order flow observable $R(\vft)$, the $R(\ac)$ shows stronger dependence on nuclear deformation with smaller statistical uncertainties, indicate that $R(\ac)$ is statistical friendly observable which are very important for model study and data analysis.

In fact, $\ac$ is largely influenced by individual flow harmonics, while the correspondent normalized asymmetric cumulant $\nac$ could reduce such flow contributions. $\nac$ directly reflect the correlation between second and fourth order event-plane $\phitf$, after neglecting the correlations between different flow harmonics. In Fig.~(\ref{fig:nac}) (a) and (b), we plot the ratios of normalized asymmetric cumulant $R(\nac)$ and $R(\phitf)$  with $\Phi_{2} = (1/2)\arctan{({\rm Im}Q_{2}/{\rm Re}Q_2)}$ and $\Phi_{4} = (1/4)\arctan{({\rm Im}Q_{4}/{\rm Re}Q_4)}$.
The $R(\nac)$ depends on $\betaRu$ and $\betaZr$ with large statistical uncertainties which is inherited from $\vft$ and $\vtf$ shown in Fig. \ref{fig:flow}.
We found the event-plane correlations ratio $R(\phitf)$  also depend on $\betaRu$ and $\betaZr$, which show similar  trend as $R(\ac)$ and $R(\vtt)$ in Fig.~\ref{fig:ac} and Fig.~\ref{fig:flow}(a). We note that the event plane correlations can also be calculated by~\cite{Bhalerao:2013ina}
\begin{equation}
c\{2,2,-4\} \equiv  \frac{\mean{Q_{2A}^{2}Q_{4B}^{*}}}{\sqrt{\mean{Q_{4A}Q_{4B}^{*}}}\sqrt{\mean{Q_{2A}^{2}Q_{2B}^{*2}}}}
\end{equation}
with the two sub-event method, and we have checked that the $R(c\{2,2,-4\})$ consist with $R(\nac)$ and $R(\phitf)$ within large statistical errors.

\begin{figure*}[!hbt]
	\begin{centering}
		\includegraphics[scale=0.32]{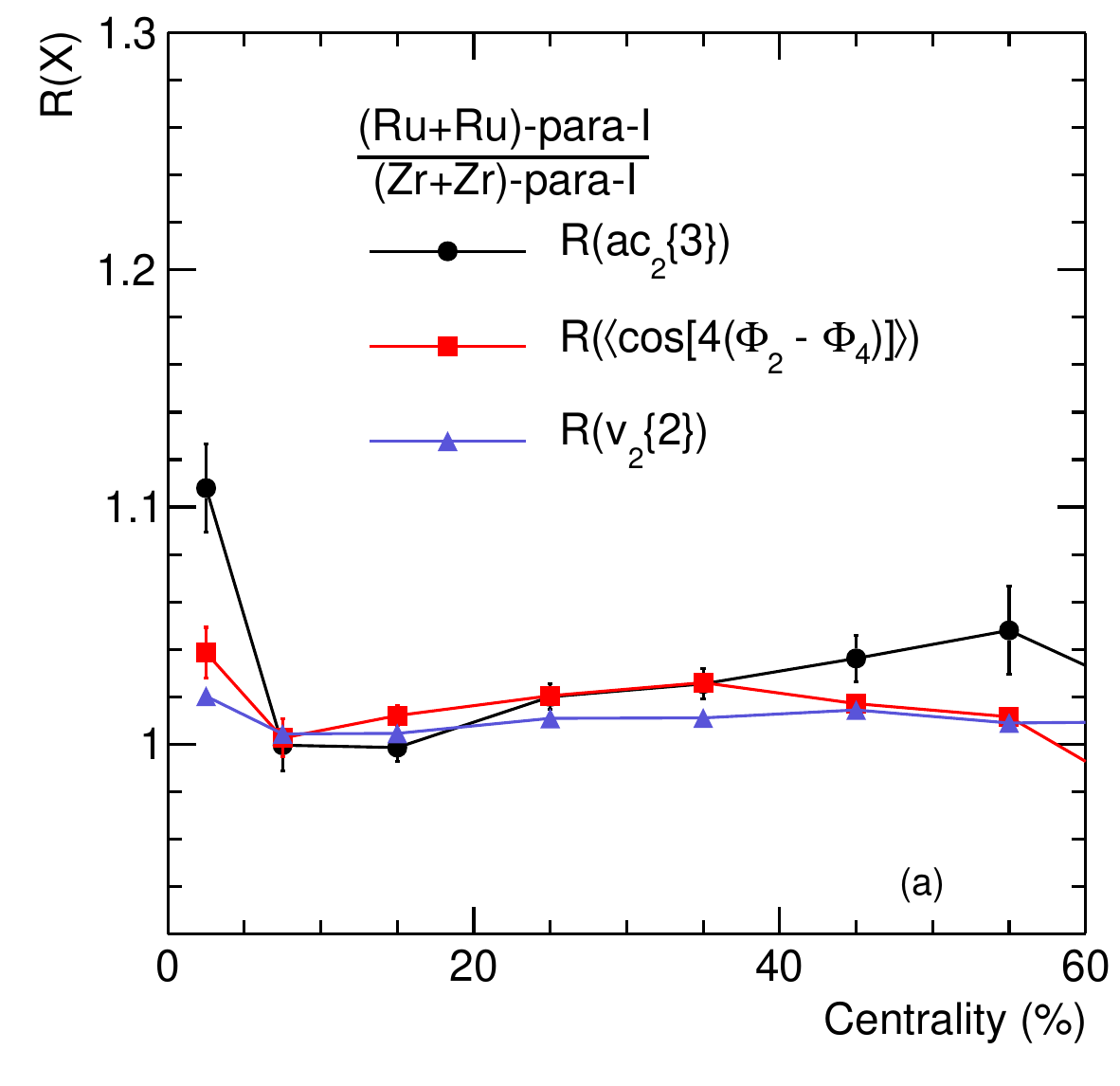}
		\includegraphics[scale=0.32]{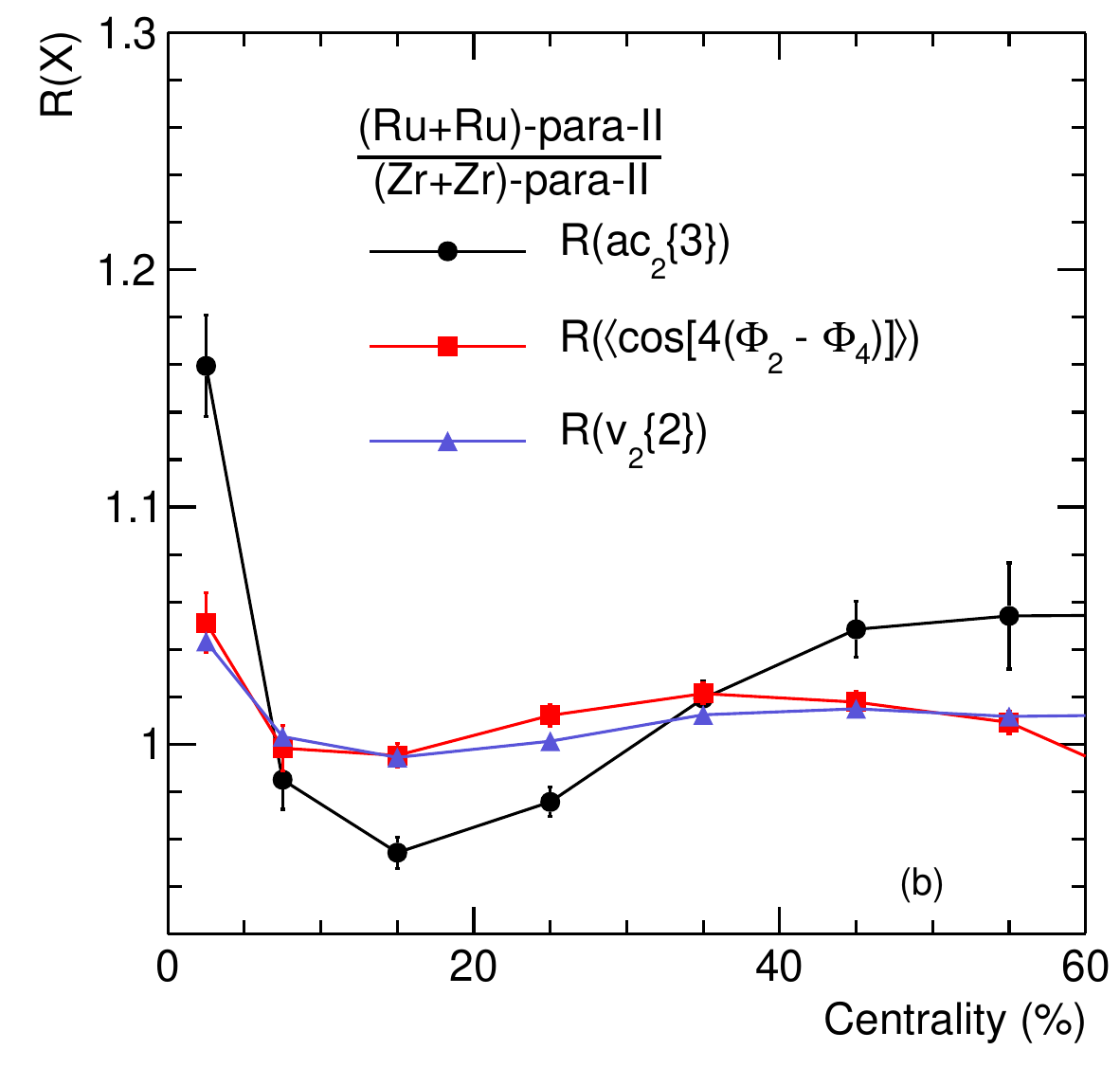}
	\par\end{centering}
	\caption{The comparison among $R(\ac)$, $R(\phitf)$ and $R(\vtt)$ at relativistic isobar collisions with different set of nuclear deformations: (a) (Ru-para-I, Zr-para-I) and (b) (Ru-para-II, Zr-para-II). The observables are calculated by the standard Q-cumulant method with $0.2<\pt<2$ \gevc\ and $|\eta|<2$. \label{fig:comp} }
\end{figure*}

The similar trends for $R(\ac)$, $R(\phitf)$, and $R(\vtt)$ indicate that both the nuclear deformation and the resulting fluctuations are important to understand the observed flow differences in isobar collisions.
In Fig.~\ref{fig:comp}, as a summary, we compare those ratios with two sets of deformations, i.e., (Ru-para-I, Zr-para-I) v.s. (Ru-para-II, Zr-para-II).
We find that the $R(\ac)$, $R(\phitf)$, and $R(\vtt)$ show different response to the deformation.
We expected that our proposed observables, together with other observables like $R(v_{3})$, can be used to constrain the initial deformation and fluctuations for relativistic isobar collisions.

Note that the non-flow contributions have not been fully included in our study, since non-flow contribution from  iEBE-VISHNU simulations are mainly from resonance decay. If the non-flow contributions are the same for the two colliding system, the observed differences should be diluted to some extend.
We have checked that the $|\eta|$<2 used in this study make the $R(\vtt)$ a little bit larger deviate from unity than the one using a smaller pseudorapidity cut $|\eta|$<1.
We also find that the sub-event method (e.g. $\Delta\eta$>0.4) can also suppress the non-flow effect, which make the $R(\vtt)$ further deviate from unity.
These effects are considerably small on $R(\vtt)$, and even not visible on other observables, partly due to large statistical uncertainties.
For the three particle cumulant $\ac$, the three sub-event method can largely suppress the non-flow contributions but restricted by statistics. Besides resonance decay, further studies with more non-flow effects included  are needed.
The dataset collected in the experiment (about 2 billion events for each collision system)~\cite{STAR:2021mii} is $14$ times larger than the model study used in this work (about $3 {\rm million}\ ({\rm hydro}) \times 50 ({\rm UrQMD\ oversamplings})= 150$ million events for each collision system),
with which observables are expected to be measured more precisely with
various non-flow subtraction methods.
The comparison between model study and the experiment data can provide more insights on the non-flow contributions.

\begin{figure}[!hbt]
	\begin{centering}
		\includegraphics[scale=0.35]{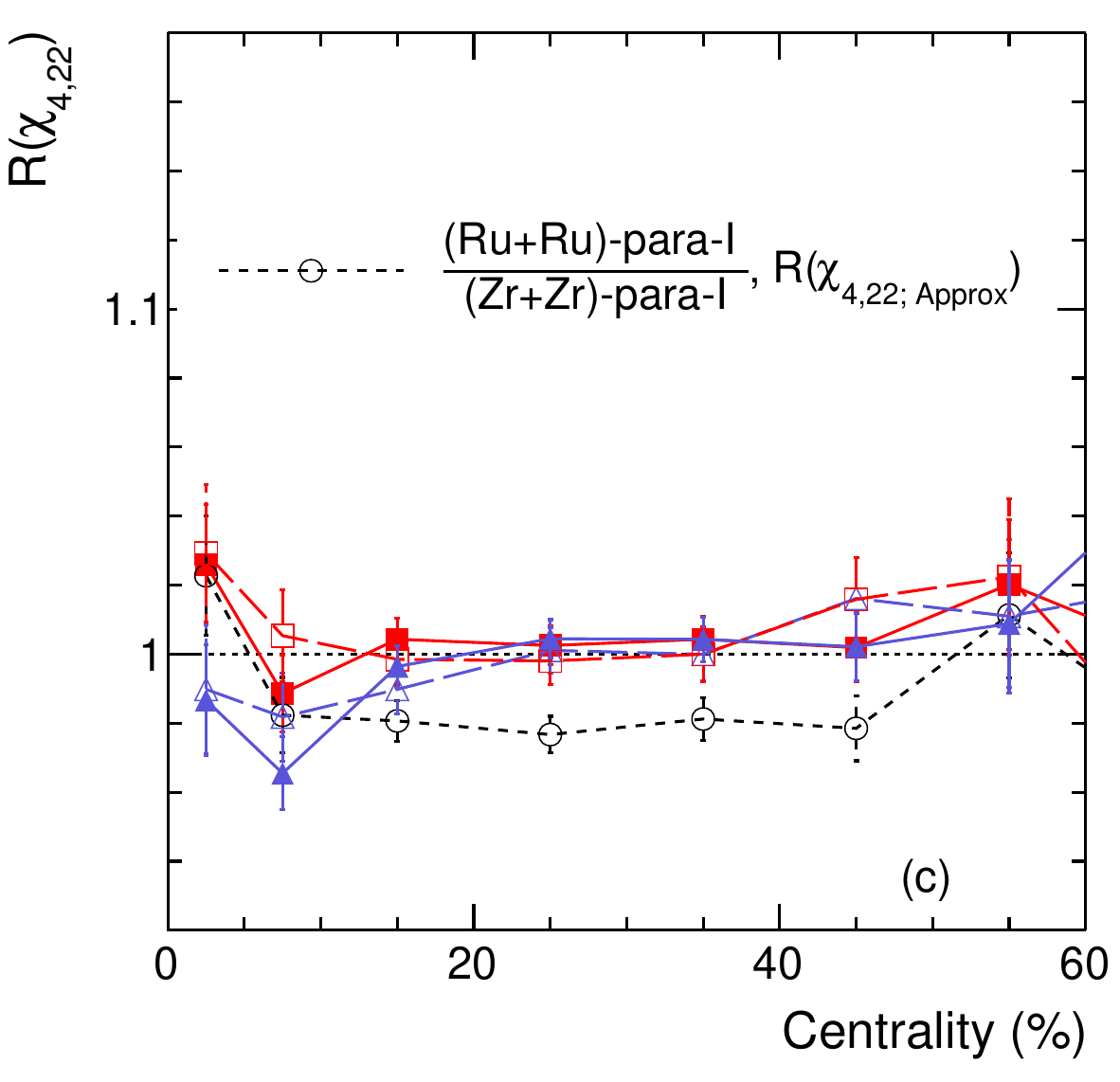}
	\par\end{centering}
	\caption{The centrality dependent $R(\chi_{4,22})$ calculated by iEBE-VISHNU model with different sets of nuclear deformations. The approximated $R(\chi)$ calculated by Eq.~(\ref{eq:approx}) is shown as open black circles.  The observables are calculated by the standard Q-cumulant method with $0.2<\pt<2$ \gevc\ and $|\eta|<2$.\label{fig:chi}}
\end{figure}

Before the end of this study, it is also interested to study the nonlinear response coefficient $\chi_{4,22}$ ratios in isobar collisions, which is defined as~\cite{Yan:2015jma}:
\begin{equation}
\chi_{4,22} \equiv \frac{\ac}{\langle v_{2}^{4}\rangle} = \nac\sqrt{\frac{\vft^{2}}{2\vtt^{4}-\vtf^{4}}}.
\end{equation}
The results are shown in Fig.~\ref{fig:chi}, which are consistent with unity within errors. It indicates that the nonlinear coefficient $\chi_{4,22}$ is not sensitive to the nuclear deformation, although the top $5\%$ results show some weak sensitivities with large uncertainties. Note that earlier study also found that the nonlinear coefficients are not sensitive to impact parameter and initial model, but mostly determined by the freezeout temperature in hydrodynamic simulation~\cite{Yan:2015jma,Luzum:2010ae}.

In the most central collisions, the $\vtf$ is significantly smaller than $\vtt$, then $\chi_{4,22}$  can be approximately expressed as:~\footnote{We thank G. Giacalone for valuable discussion on this point.}
\begin{equation}
\chi_{\rm 4,22;Approx} = \frac{\ac}{\vtt^{4}}.\label{eq:approx}
\end{equation}
Fig.~\ref{fig:chi} also shows $R(\chi_{\rm 4,22;Approx})$, calculated with $\betaRu=0.12$ and $\betaZr=0.16$, with open black circles. We find that this approximation works well for top $10\%$ centrality, but failed at semi-central and peripheral collisions. Such deviation at non-central collisions indicates that fluctuations are essential to understand the flow differences in relativistic isobar collisions.

\section{Summary}
The observed differences between flow harmonics for \RuRu\ and \ZrZr\ collisions at $\snn=200$ GeV provide unique opportunities to probe the nuclear structure of the colliding nuclei.
In this letter, we proposed that the asymmetric cumulant ratio $R(\ac)$, together with the corresponding individual flow harmonic ratios $R(v_{n})$ and event-plane correlation ratio $R(\nac)$ ($R(\phitf)$), can simultaneously constrain the nuclear deformation and the resulting fluctuations. Our iEBE-VISHNU hybrid model simulations indicate that the statistical friendly observable $R(\ac)$ is very sensitive to the  quadrupole and octupole deformation of $\beta_2$ and  $\beta_3$. To further investigate this sensitivity, we divided the $\ac$ into three parts, i.e., $v_2$ ($\vtt$ and $\vtf$), $v_{4}$ ($\vft$), and the normalized asymmetric cumulants $\nac$, but ignore their correlations and non-flow effect. We found that both the flow harmonics differences and event-plane correlation differences in the isobar collisions depend on $\betaRu$ and $\betaZr$.
The event-plane correlation differences on the nuclear structure could be larger than the elliptic flow difference, indicating the importance of initial fluctuations. The $R(\ac)\simeq R(\vtt^{4})$ works well in the most central collisions, but show obvious deviation in non-central collisions. We found the nonlinear coefficients extracted from the $\ac$ are identical in the two systems, indicating the insensitivity of $\chi_{4,22}$ to the initial state and the details of nuclear deformation.

\section*{Acknowledgments}
 We thank G.~Giacalone, J.~ Jia, F.~ Wang and W. Zhao for useful discussions.
We thank Yu Hu for providing the STAR data.
 This work is supported in part by the National Natural Science Foundation of China (Grant Nos.12075007, 12275082, 12035006, 12075085, 11905059),
and the Zhejiang Provincial Natural Science Foundation of China (Grant No.~LY21A050001).

\bibliographystyle{elsarticle-num}
\bibliography{ref2}
\end{document}